# High-temperature intrinsic ferromagnetism in heavily Fe-doped GaAs layers

A.V. Kudrin[1,*], V.P. Lesnikov[1], Yu.A. Danilov[1], M.V. Dorokhin[1], O.V. Vikhrova[1], P.B. Demina[1], D.A. Pavlov[1], Yu.V. Usov[1], V.E. Milin[1], Yu.M. Kuznetsov[1], R.N. Kriukov[1], A.A. Konakov[1] and N.Yu. Tabachkova [2,3]

[1] *Lobachevsky State University of Nizhny Novgorod, Gagarin av. 23/3, 603950 Nizhny Novgorod, Russia*
[2] *National University of Science and Technology "MISiS", 119049 Moscow, Russia*
[3] *Prokhorov General Physics Institute of the Russian Academy of Sciences*

*kudrin@nifti.unn.ru

The layers of a high-temperature novel GaAs:Fe diluted magnetic semiconductor (DMS) with an average Fe content up to 20 at. % were grown on (001) *i*-GaAs substrates using a pulsed laser deposition in a vacuum. The transmission electron microscopy (TEM) and energy-dispersive X-ray spectroscopy investigations revealed that the conductive layers obtained at 180 and 200 ºC are epitaxial, do not contain any second-phase inclusions, but contain the Fe-enriched columnar regions of overlapped microtwins. The TEM investigations of the non-conductive layer obtained at 250 ºC revealed the embedded coherent Fe-rich clusters of GaAs:Fe DMS. The X-ray photoelectron spectroscopy investigations showed that Fe atoms form chemical bonds with Ga and As atoms with almost equal probability and thus the comparable number of Fe atoms substitute on Ga and As sites. The *n*-type conductivity of the obtained conductive GaAs:Fe layers is apparently associated with electron transport in a Fe acceptor impurity band within the GaAs band gap. A hysteretic negative magnetoresistance was observed in the conductive layers up to room temperature. Magnetoresistance measurements point to the out-of-plane magnetic anisotropy of the conductive GaAs:Fe layers related to the presence of the columnar regions. The studies of the magnetic circular dichroism confirm that the layers obtained at 180, 200 and 250 ºC are intrinsic ferromagnetic semiconductors and the Curie point can reach up to at least room temperature in case of the conductive layer obtained at 200 ºC. It was suggested that in heavily Fe-doped GaAs layers the ferromagnetism is related to the Zener double exchange between Fe atoms with different valence states via an intermediate As and Ga atom.

## 1. Introduction

Heavily Fe-doped diluted magnetic semiconductors (DMSs) are considered as promising materials for semiconductor spintronics since the recent synthesis of the (Ga,Fe)Sb [1-4] and (In,Fe)Sb [5,6] DMSs with a Fe content up to 17 at. % and above-room temperature (RT) Curie point ($T_C$). At present, the nature of ferromagnetism in Fe doped narrow-gap ((In,Fe)As [7], (Ga,Fe)Sb, (In,Fe)Sb) and wide-gap III-V ((Al,Fe)Sb [8]) semiconductors is a subject of active research.

Apparently the ferromagnetism in (III,Fe)V DMSs is not carrier-mediated [5-8]. In particular, in the ferromagnetic layers of wide-gap (Al,Fe)Sb DMS with $T_C$ = 40 K the carrier concentration is about $10^{17}$ cm$^{-3}$ [8]. In the layers of narrow-gap (In,Fe)Sb DMS with Curie point above RT the carrier density is about $10^{17}$ - $10^{18}$ cm$^{-3}$ [5,6]. Such low concentrations should not provide the exchange interaction between magnetic atoms via charge carriers. One can compare the ferromagnetic Mn doped III-V DMSs in which the carrier concentration is typically about $10^{20}$ - $10^{21}$ cm$^{-3}$, while $T_C$ does not exceed 200 K [9,10]. In [8,11] it was assumed that the ferromagnetism of (III,Fe)V DMSs (in particular in (Al,Fe)Sb and (In,Fe)Sb) is associated with the ferromagnetic coupling between the second nearest neighbor Fe atoms via the super-exchange interaction. In [12] we revealed a weak interrelation between the ferromagnetism and the charge carrier concentration in (In,Fe)Sb that is fundamentally different from Mn doped III-V DMS.

Obtained results raise the question of creating the ferromagnetic layers of a practically important GaAs semiconductor by introducing a large amount of Fe atoms (> 10 at. %) during an epitaxial growth.

There are a few papers from the early 2000s about the formation of the (Ga,Fe)As layers by molecular beam epitaxy (MBE) with a relatively low Fe content (up to 3.5 at. %). In [13-15] the Ga$_{1-x}$Fe$_x$As layers with *x* in the range of 0.007 – 0.07 were grown at the temperatures in the range of 260 – 580 ºC. These layers were high-resistive with a weak *n*-type conductivity. The structures obtained at 260 ºC were characterized as Ga$_{1-x}$Fe$_x$As homogeneous paramagnetic epilayers, whereas the structures obtained at 350 – 580 ºC were GaAs epilayers with second-phase ferromagnetic inclusions. These (Ga,Fe)As layers were obtained at the relatively high temperatures and contained the relatively low amount of iron. In this work we present the results of investigations of GaAs:Fe layers with high Fe content (the average Fe content up to 20 at. %) obtained at 180 – 350 ºC, including conductive layers with the ferromagnetic properties up to room temperature.

## 2. Material and methods

The GaAs and GaAs:Fe layers were grown by a pulsed laser deposition (PLD) in a vacuum on semi-insulating (001) GaAs substrates [2,5,12,17]. The Fe content was set by the technological parameter $Y_{Fe} = t_{Fe}/(t_{Fe}+t_{GaAs})$, where $t_{Fe}$ and $t_{GaAs}$ are the ablation times of the Fe and GaAs targets, respectively. The $Y_{Fe}$ was varied in the rage of 0 – 0.25. The substrate temperature $T_g$ was varied in the rage of 180 – 350 ºC. The nominal thickness of the GaAs:Fe layers was ~ 40 nm.

The surface of the structures was examined by atomic force microscopy (AFM). The structural properties of the samples were investigated by an analytical transmission electron microscopy (TEM). The distribution of constituent elements was obtained by an energy-dispersive X-ray spectroscopy (EDS) during microscopy investigations.

The chemical composition of the structures was also studied by X-ray photoelectron spectroscopy (XPS). To excite photoemission, the Al Kα emission line with the energy of 1486.7 eV was used. The diameter of the analysis region was 3 mm. The profiling of the composition over the depth of the sample was carried out by a surface etching with the 1 kV Ar$^+$ beam. The concentration of chemical elements was determined by the method of relative sensitivity factors (RSF) [16]. During the XPS measurements the As 3$d$, Ga 3$d$, Fe 2$p_{3/2}$, O 1$s$ and C 1$s$ photoelectron lines were recorded.

Optical reflectivity spectra were obtained at RT in the spectral range of 2.4 – 6 eV. The $dc$ magnetotransport measurements were carried out in the van der Pauw and Hall bar geometry in a closed-cycle He cryostat. The investigations of the magnetic circular dichroism (MCD) were performed in the cryostat in the spectral range from 1.37 to 2.03 eV using a monochromatized emission of a xenon arc lamp as a light source.

In this study we present results for the structures are listed in Table I.

Table I. Studied GaAs:Fe samples. $T_g$ –growth temperature (ºC). $Y_{Fe}$ – technological Fe content, Fe content from EDS investigations, $T_C$ from magnetotransport and MCD measurements.

| Structure | $T_g$ (ºC) | $Y_{Fe}$ | Fe content (at. %) | $T_C$ (K) |
|---|---|---|---|---|
| 200-0 | 200 | 0 | - | - |
| 200-8 | 200 | 0.08 | 6 | - |
| 200-17 | 200 | 0.17 | 14 | ~ 200 |
| 200-25 | 200 | 0.25 | 20 | ~ 170 |
| 180-25 | 180 | 0.25 | 20 | above 295 |
| 250-25 | 250 | 0.25 | 20 | - |
| 350-25 | 350 | 0.25 | 20 | - |

## 3. Results

*3.1. Structural and optical properties*

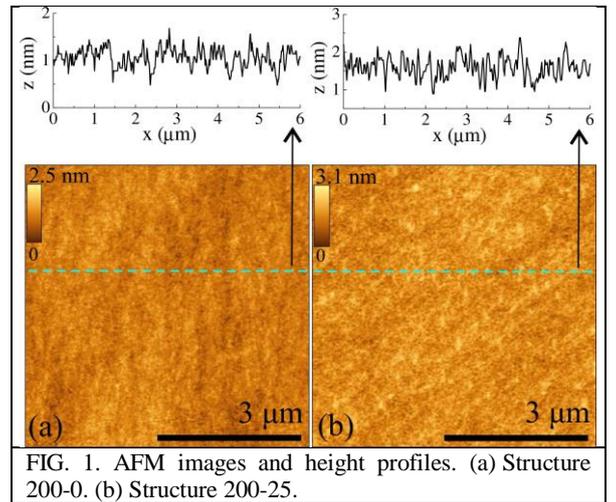

FIG. 1. AFM images and height profiles. (a) Structure 200-0. (b) Structure 200-25.

Figure 1 shows the AFM surface morphology of the structures 200-0 (Figure 1(a)) and 200-25 (Figure 1(b)). All obtained GaAs:Fe layers (and the GaAs layer) have the smooth surface with a root mean square (RMS) roughness of about 0.3 nm.

Figure 2(a) shows the overview high-resolution TEM (HRTEM) image of a 95 nm long region for the GaAs:Fe/GaAs structure 180-25. The image demonstrates a smooth homogeneous epitaxial GaAs:Fe layer with the thickness of about 45 nm without evident crystalline second-phase inclusions which can usually appear in (III,Fe)V structures in the form of a spherical or oval clusters with a moiré-type contrast [15,17]. The TEM images reveal a large number of stacking faults and microtwins along {111} planes, that appear as the assemblage of straight lines at an angle of ≈ 70º with respect to each other (Figure 2(a)). The formation of stacking faults and microtwins in epitaxial layers typically is a consequence of the lattice mismatch between an epilayer and a substrate (or a buffer layer) [5,18]. In particular for GaAs layers, similar V-shape stacking defects were observed in the GaAs/Si [19], Ga(As,Bi)/GaAs [20,21] structures and even in the low temperature (LT) GaAs epilayers grown on GaAs substrate (since the lattice parameter of the as-grown LT-GaAs layer is larger than that of the bulk GaAs by 0.1%) [22]. The inset of Figure 2(a) shows a fast Fourier transform (FFT) diffraction pattern of the HRTEM image. The FFT diffraction pattern contains bright primary diffraction spots corresponding to a zinc-blende type lattice of the epitaxial GaAs:Fe matrix and additional spots between primary spots. The observation of two additional weak equidistant spots between the pairs of primary spots reveal the presence of additional triple periodicity in the GaAs:Fe layer.

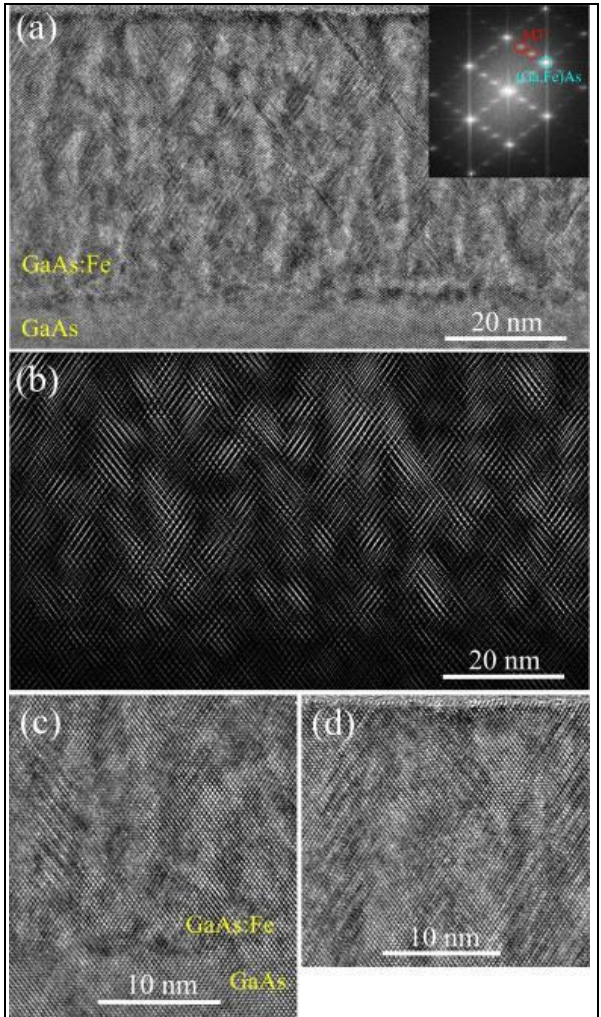

FIG. 2. Results of cross-section TEM investigations of GaAs:Fe/GaAs structure 180-25. (a) Overview HRTEM image of 95 nm long region (inset shows FFT diffraction pattern of image). (b) Inverse FFT image of GaAs:Fe layer obtained from microtwins (MT) spots of FFT diffraction pattern. (c) HRTEM image of 24×27 nm$^2$ area near substrate, (d) HRTEM image of 24×23 nm$^2$ area near surface.

Similar triple periodicity was observed in the GaAs/Si, LT-GaAs/GaAs, Ga(As,Bi)/GaAs structures and was explained by the presence of the regions of the overlapped twins [19-22]. The Figure 2(b) exhibits the inverse FFT image of the GaAs:Fe layer obtained from the additional spots of the FFT diffraction pattern. The inverse FFT image reveals, that the regions of the overlapped twins (triple periodicity regions) form periodic columnar-shape domains (Figure 2(b)). For GaSb and InSb matrices the introduction of large number of Fe atoms leads to the lattice constant decrease [1,6]. Since the Fe atomic radius is less than the Ga and As atomic radii, the GaAs:Fe lattice constant should also be smaller than the GaAs lattice constant. Apparently this is the cause of the microtwins (and consequently triple periodicity) regions appearance in the obtained GaAs:Fe layer. It can be assumed that these regions have the differing lattice constant due to the higher concentration of Fe atoms. In particular, for the Ga(As,Bi) layers it was observed, that the triple periodicity regions have the higher Bi content [20,21]. The Figures 2(c) and 2(d) show the close-up HRTEM images of the GaAs:Fe layer near the GaAs substrate (the 24×27 nm$^2$ area, Fig. 2(c)) and near the surface (the 24×23 nm$^2$ area, Fig. 2(d)). These HRTEM images confirm the epitaxial growth character of the GaAs:Fe layer, the absence of crystalline second-phase inclusions and the presence of the V-shape domains of the microtwins that originate at the GaAs:Fe/GaAs interface and propagate further toward the surface.

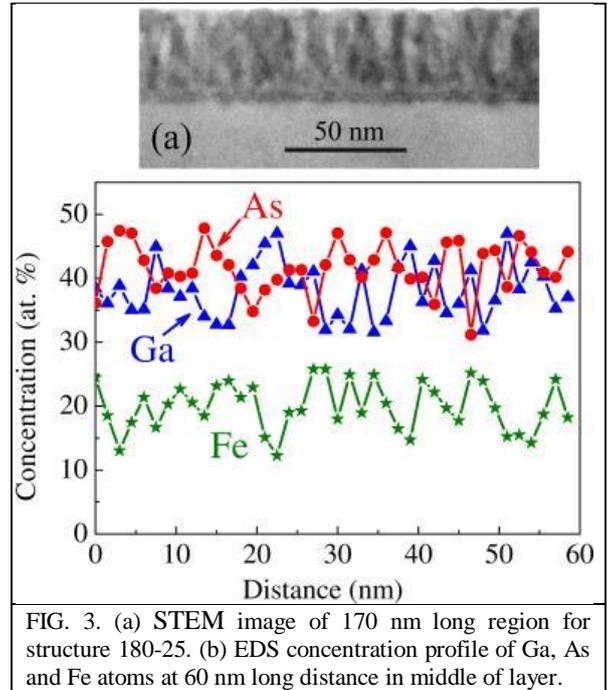

FIG. 3. (a) STEM image of 170 nm long region for structure 180-25. (b) EDS concentration profile of Ga, As and Fe atoms at 60 nm long distance in middle of layer.

Figure 3(a) presents a scanning cross-section TEM (STEM) image of the structure 180-25. The STEM image also reveals the presence of periodic columnar-shape structure with the different contrast due to the fluctuations of constituent elements across the periodic columnar domains of overlapped microtwins. The latter can probably be attributed to some fluctuations in the local Fe concentration. The similar structure with the nano-columnar Fe-enriched DMS regions was observed in the 40-nm-thick (Ga$_{0.7}$,Fe$_{0.3}$)Sb layer obtained by MBE on the AlSb buffer layer [4]. Figure 3(b) presents the EDS concentration profile of Ga, As and Fe atoms at 60 nm long distance in middle of the GaAs:Fe layer for the structure 180-25. The modulation of the concentration of Fe atoms is observed which is consistent with HRTEM and STEM data. The EDS investigations show what the average Fe content in the GaAs:Fe layer for the structure 180-25 is about 20 at. %. The results also reveal the comparable concentration of Ga and As atoms in the layer which points to that Fe atoms replace not only Ga but also As atoms.

Figure 4(a) shows the overview HRTEM image of a 73 nm long region for the GaAs:Fe/GaAs structure 200-25. As for the structure 180-25 (Fig. 2), the GaAs:Fe layer is a smooth homogeneous epitaxial layer with the thickness of about 35 nm without evident crystalline second-phase inclusions. The GaAs:Fe layer of the structure 200-25 also contains the regions of the overlapped twins (triple periodicity regions, Fig 4(a) and the inset of Fig. 4(a)). The inverse FFT image of the GaAs:Fe layer obtained from the additional spots of the FFT diffraction pattern (the inset of Fig. 4(a)) reveal, that the overlapped microtwins also form columnar-shape domains and these columnar-shape domains are wider (Fig. 4(b)) than that for the structure 180-25 (Fig. 2(b)). The close-up HRTEM images of the 15×17 nm$^2$ area near the GaAs substrate (Fig. 4(c)) and of the 15×11 nm$^2$ area near the surface (Fig. 4(d)) as for the structure 180-25 (Fig. 2) confirm the epitaxial growth character of the GaAs:Fe layer, the absence of crystalline second-phase inclusions and the presence of the domains of microtwins that propagate from the substrate to the surface.

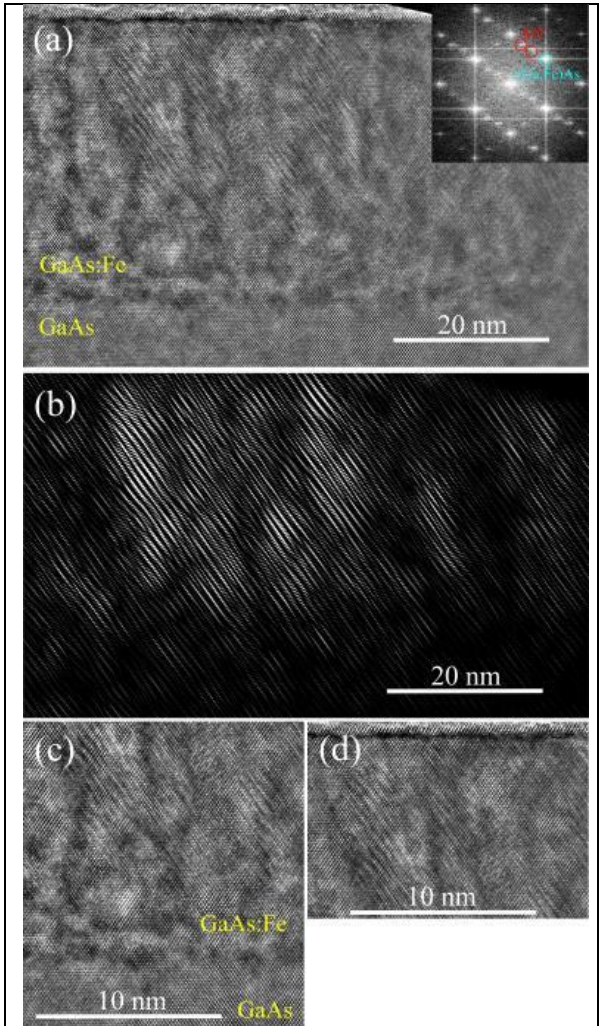

FIG. 4. Results of cross-section TEM investigations of GaAs:Fe/GaAs structure 200-25. (a) Overview HRTEM image of 73 nm long region (inset shows FFT diffraction pattern of image). (b) Inverse FFT image of GaAs:Fe layer obtained from microtwins (MT) spots of FFT diffraction pattern. (c) HRTEM image of 15×17 nm$^2$ area near substrate, (d) HRTEM image of 15×11 nm$^2$ area near surface.

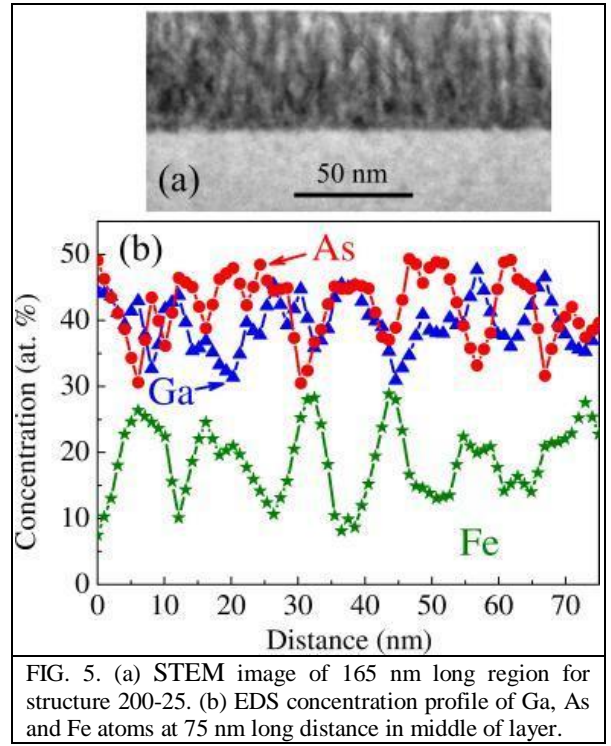

FIG. 5. (a) STEM image of 165 nm long region for structure 200-25. (b) EDS concentration profile of Ga, As and Fe atoms at 75 nm long distance in middle of layer.

Figure 5(a) presents the STEM image of the structure 200-25. The STEM image confirms the presence of periodic columnar-shape structure with the different contrast which is again attributed to the fluctuations of constituent elements including the local Fe concentration.

Figure 5(b) presents the EDS concentration profile of Ga, As and Fe atoms at 75 nm long distance in middle of the GaAs:Fe layer for the structure 200-25. The elements profiling reveals the clear modulation of the Fe content with a period consistent with the period of the columnar-shape domains on HRTEM images (Fig. 4). The EDS data support the assumption that the periodic columnar-shape regions formed by the overlapping twins have the higher Fe concentration than the regions between them. For the structure 200-25 these columnar-shape regions are wider (Fig. 4(b)) and have the higher local Fe concentration (Fig. 5(b)) than that of the structure 180-25 (Fig. 2(b) and (Fig. 3(b))) which can be explained by the faster diffusion of Fe atoms at the higher growth temperature. The presence of columnar Fe-enriched GaAs:Fe regions separated by GaAs:Fe regions with a lower Fe concentration should effect the transport and magnetic properties of the obtained layers. Note that the Fe concentration in the columnar Fe-enriched GaAs:Fe regions is up to 28 at. % (the method error is ± 4 at. %), hence these regions cannot be characterized as regions of the known Fe$_3$GaAs phase (with a Fe content of 60 at. %) that forms in the high temperature Fe-doped GaAs [23]. For the structure 200-25 as for the structure 180-25 the EDS investigations also reveal approximately equal content of Ga and As atoms in the layer. The EDS investigations show what the average Fe content in the GaAs:Fe layer for the structure 200-25 is about 20 at. % as for the structure 180-25. Taking into account this average Fe concentration for the structures 180-25, 200-25 ($Y_{Fe} = 0.25$) and the technological Fe content for the structures 200-17 ($Y_{Fe} = 0.17$) and 200-8 ($Y_{Fe} = 0.08$), the average Fe concentration can be estimated as about 14 and 6 at. % for the structure 200-17 and 200-8, respectively.

An increase in the temperature of the growth process to 250 ºC leads to a significant change in the crystal structure of the GaAs:Fe layer. Figure 6(a) shows the overview TEM image from a 73 nm long region for the GaAs:Fe/GaAs structure 250-25. In contrast to the structures 180–25 and 200–25 (Figures 2(a) and 4(a)), the presence of elongated clusters oriented mainly along the growth direction is observed in the GaAs:Fe layer of the structure 250–25 (Fig. 6(a)).

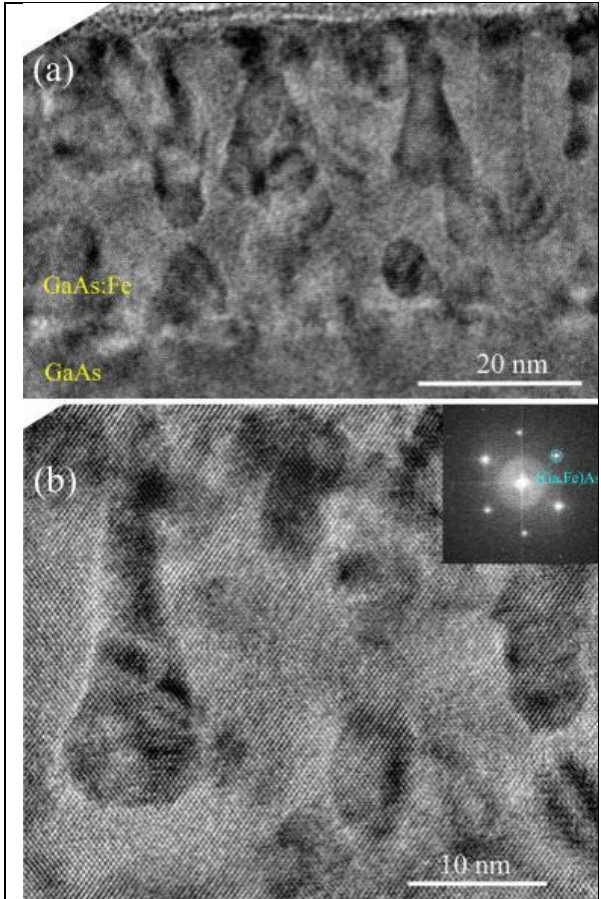

FIG. 6. Results of cross-section TEM investigations of GaAs:Fe/GaAs structure 250-25. (a) Overview TEM image of 71 nm long region. (b) HRTEM image of 42×36 nm² area (inset shows FFT diffraction pattern of image)

The Figure 6(b) shows HRTEM images of the 42×36 nm² area with clusters for the structure 250-25. The HRTEM reveal that the clusters have zinc-blende crystal structure and coherently embedded in GaAs:Fe matrix without clear misfit dislocation. The FFT diffraction pattern of the HRTEM image also contains only diffraction spots corresponding to a zinc-blende type lattice of the epitaxial GaAs:Fe (the inset of Fig. 6(b)). It is obvious, that the formation of the clusters is associated with the diffusion rate increase of Fe atoms with increasing growth temperature. For the case of $T_g = 250$ ºC the formation of zinc-blende cluster-like regions with the higher local Fe concentration occurs but not the formation of metallic Fe or some intermetallic second-phase clusters. Thus, the GaAs:Fe layer for the structure 250-90 can be defined as the GaAs:Fe DMS matrix with the embedded coherent Fe-rich clusters of GaAs:Fe DMS. It can also be concluded that the growth temperature increase leads to the evolution of the columnar Fe-enriched domains of microtwins (Figures 2 and 4) to the coherent elongated Fe-rich clusters of GaAs:Fe DMS (Fig. 6). The concentration of Fe-rich clusters of GaAs:Fe DMS is below the percolation threshold and this should has the influence on the conductivity of the structure 250-90.

Figure 7 shows the XPS depth distribution profiles of C, O, Ga, As and Fe atoms for the sample 180-25. The presence of oxygen atoms on the surface is related to the oxidation in air. The average Fe content in the GaAs:Fe layer is $19 \pm 1$ at. %. Throughout the profile, the concentration of Ga atoms is approximately equal to the concentration of As atoms which is consistent with EDS data (Figures 3 and 5).

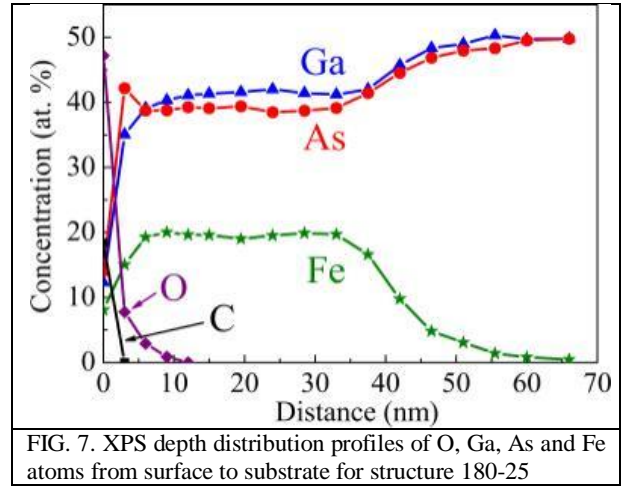

FIG. 7. XPS depth distribution profiles of O, Ga, As and Fe atoms from surface to substrate for structure 180-25

Using the method described in Ref. [24], the As 3$d$, Ga 3$d$ and Fe 2$p_{3/2}$ photoelectron lines (obtained during the profiling) were analyzed. The analysis of the As 3$d$ and Ga 3$d$ photoelectron lines showed that in the GaAs:Fe layer the Ga-As, As-Fe and Ga-Fe chemical bonds are present, and in the GaAs substrate only the Ga-As bonds are present. The identification of the spectral components was carried out using the data from [25].

The analysis of the Fe 2$p_{3/2}$ photoelectron line did not reveal the presence of Fe-Fe bond (within the error of the method about 1 at.%), but the presence of Fe-As and Fe-Ga bonds was detected. It is not possible to reliably separate Fe-As and Fe-Ga bonds within the method used, only the summary concentration of Fe-As and Fe-Ga chemical bonds can be detected.

Figure 8 shows for the structure 180-25 the depth profile of the concentration of the following chemical bonds: Ga atoms bonded to As atoms, As atoms bonded to Ga atoms, Ga atoms bonded to Fe atoms, As atoms bonded to Fe atoms, all chemically bonded Fe atoms (Fe-Ga, Fe-As).

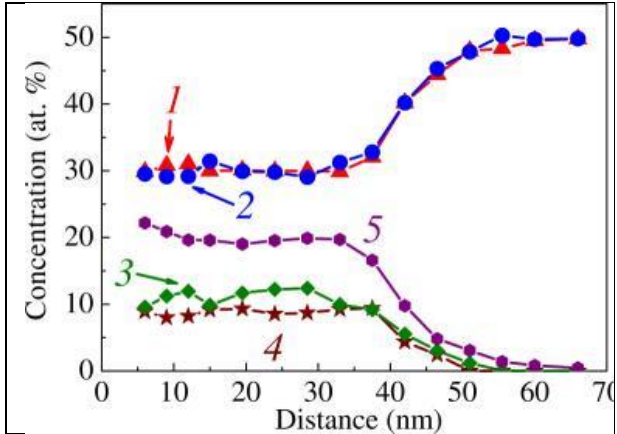

FIG. 8. Depth profile of concentration of atoms with different chemical bonds for structure 180-25: *1* - Ga atoms bonded to As atoms, *2* - As atoms bonded to Ga atoms, *3* - Ga atoms bonded to Fe atoms, *4* - As atoms bonded to Fe atoms, *5* - all chemically bonded Fe atoms (Fe-Ga, Fe-As).

The results of the analysis of chemical bonds showed that there is a comparable number of Ga and As atoms chemically bonded to Fe atoms, therefore, the comparable number of Fe atoms substitute on Ga and As sites. The absence of the Fe-Fe chemical bonds indicates that the GaAs:Fe layer does not contain metallic iron clusters.

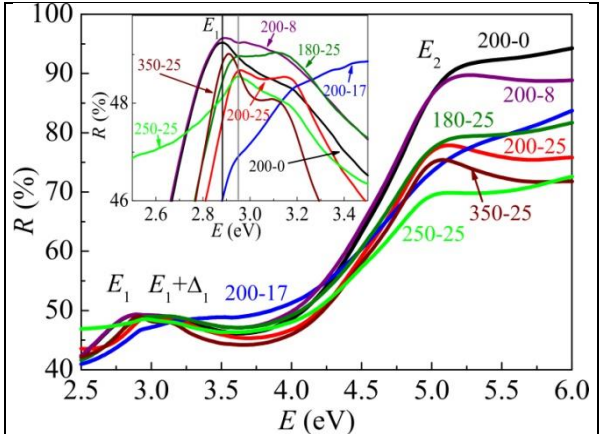

FIG. 9. Optical reflectivity spectra at 295 K for GaAs:Fe/GaAs structures. Inset shows enlarged spectra in $E_1$ region.

Figure 9 exhibits optical reflectivity spectra at 295 K for the GaAs:Fe/GaAs structures. The reflectivity spectra for the GaAs:Fe layers agree with the spectrum for the undoped GaAs layer (the structure 200-0) and contain features associated with characteristic interband transitions for a bulk GaAs [26]. In particular, the doublet in the $E_1$ region and the intense peak in the $E_2$ region are observed. This indicates the conservation of the GaAs band structure for the GaAs:Fe layers.

For the structures 200-17, 180-25, 200-25, 250-25 and 350-25 the $E_1$ peak has a clear blueshift (the inset of Fig. 7). The blueshift of the $E_1$ peak position with the Fe concentration was observed earlier for the MBE and PLD-grown (Ga,Fe)Sb [17,27] and (In,Fe)Sb layers [5,6].

*3.2. Transport and magnetic properties*

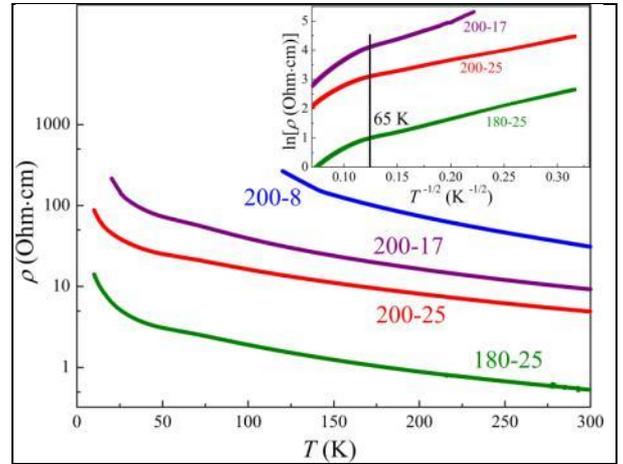

FIG. 10. Temperature dependences of resistivity for structures 200-8, 200-17, 180-25 and 200-25. The inset shows $\ln\rho$ plotted against $T^{-1/2}$ for structures 180-25, 200-17, 180-25.

Figure 10 shows the temperature dependences of a resistivity $\rho(T)$ for the structures 200-8, 200-17, 180-25 and 200-25. The resistivity of the undoped GaAs layer (the structure 200-0) is high and comparable with the resistivity of the *i*-GaAs substrate. For the conducting GaAs:Fe layers the $\rho(T)$ dependences are relatively weak. The resistivity increases by a factor of five in the temperature range of 300 - 77 K for the structures 180-25, 200-17, 180-25 and in the temperature range of 300 - 150 K for the structure 200-8. With the further temperature decrease the sharper resistivity increase is observed. The inset of Figure 10 shows $\ln\rho$ plotted against $T^{-1/2}$ for the structures 180-25, 200-17, 180-25. The close to linear resistivity dependence in these coordinates at the temperatures below about 65 K indicates that the conductivity is provided by the transfer of electrons through the electronic states localized close to the Fermi level [28]. This can be observed in lightly and heavily doped compensated semiconductors at low temperatures while the conductivity is not related to the transport of carriers in the valence or conduction bands [28].

The resistivity of the GaAs:Fe layers with $T_g = $ 200 ºC decrease with the Fe concentration increase. Note that the resistivity value at a given temperature for the structure 180-25 is much lower (about an order of magnitude) than that for the structure 200-25 (Fig. 10).

The growth temperatures increase should lead to the greater inhomogeneity of the distribution of the Fe atoms within the GaAs:Fe layer which in turn leads to the conductivity decrease. Hence for the structures 250–25 and 350–25 the resistance is comparable to the resistance of *i*-GaAs substrate.

For the structures 200-8, 200-17, 180-25 and 200-25 the Seebeck coefficient at room temperature corresponds to the *n*-type conductivity which is consistent with earlier works on the paramagnetic GaAs:Fe layers [13-15].

In general, the resistivity of GaAs:Fe layers is relatively high. The consequence of this is that the measured Hall resistance dependences on an external magnetic field ($R_{MeasH}(B)$) are completely determined by the magnetic field dependences of the longitudinal resistivity (magnetoresistance) due to the nonideal Hall measurement geometry of the samples (both for the Hall bars and van der Pauw geometry). For our samples the true Hall resistance dependences on an external magnetic field (an odd function with respect to the *B* polarity) cannot be correctly extracted from the $R_{MeasH}(B)$ dependences. The insets of Figures 11(a) and 10(a) show the $R_{MeasH}(B)$ curves at 77 and 300 K for the structures 180-25 and 200-25, respectively, for $B = \pm 1.1$ T applied perpendicularly to the sample plane. These are typical $R_{MeasH}(B)$ dependences for all our GaAs:Fe layers. In this connection, we will limit ourselves to the analysis of the magnetoresistance observed in the GaAs:Fe layers at different temperatures for the out-of-plane (*B* is applied perpendicularly to the sample plane) and in-plane (*B* is applied in the sample plane) orientations.

Figure 11(a) shows the magnetoresistance (MR = $(\rho(B) - \rho(0))/\rho(0)$) curves for the sample 180-25 at temperatures between 9 – 295 K with *B* applied perpendicular to the plane. A negative hysteretic MR is observed up to 170 K. At 9 K the out-of-plane MR curve has no clear saturation in the magnetic field of $\pm 0.36$ T (the maximum magnetic field available for investigations in the closed-cycle He cryostat). For temperatures above 9 K the out-of-plane MR curves demonstrate the saturation at $B > 0.15 – 0.2$ T (Figure 11(a). The negative MR with a clear saturation is observed up to 240 K.

Figure 11(b) shows the MR curves for the sample 180-25 at various temperatures with *B* applied in the plane. For the in-plane orientation the magnetoresistance is also negative with the hysteretic character up to 170 K. However, the shape of the MR curves at a given temperature are different for the out-of-plane and in-plane orientations. The MR measurements show that at 30 and 75 K the coercivity of the GaAs:Fe layer is higher for *B* applied perpendicular to the sample plane (Fig. 11). Also the MR investigations reveal that the magnetization of the GaAs:Fe layer of the structure 180-25 easily saturates when *B* is applied perpendicular to the sample plane (Fig. 11). These results point to the out-of-plane magnetic anisotropy of the GaAs:Fe layer which is apparently related to the presence of the columnar Fe-enriched GaAs:Fe regions. The similar out-of-plane magnetic anisotropy was observed for the MBE ($Ga_{0.7},Fe_{0.3}$)Sb layers with nano-columnar Fe-enriched DMS regions [4]. The observation of the hysteresis in the magnetoresistance up to 170 K points to the Curie temperature about 170 K for the structure 180-25.

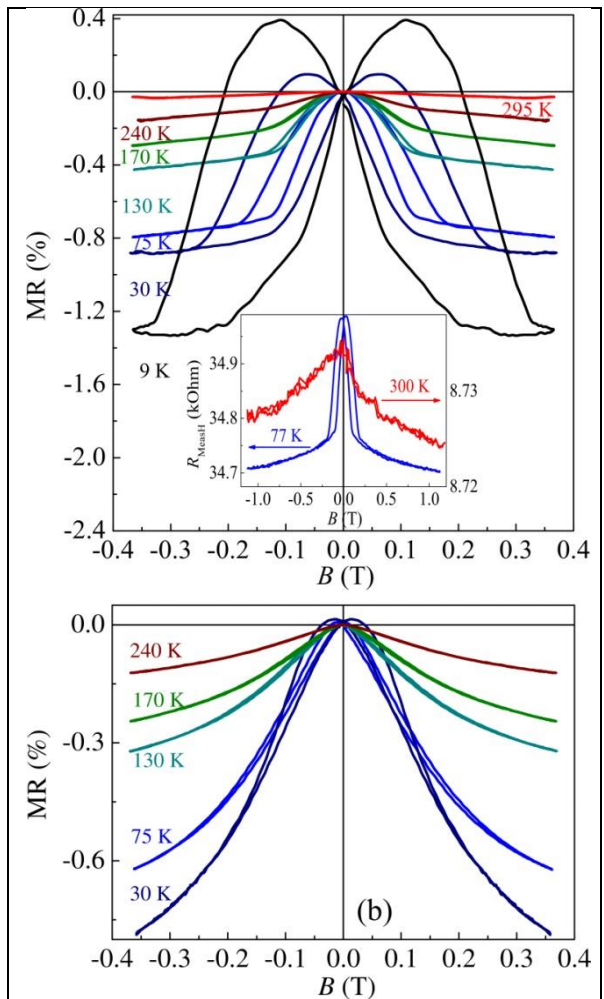

FIG. 11. MR curves at various temperatures for structure 180-25. (a) *B* is applied perpendicular to sample plane (inset shows $R_{MeasH}(B)$ curves for $B = \pm 1.1$ T at 77 and 300 K). (b) *B* is applied in sample plane.

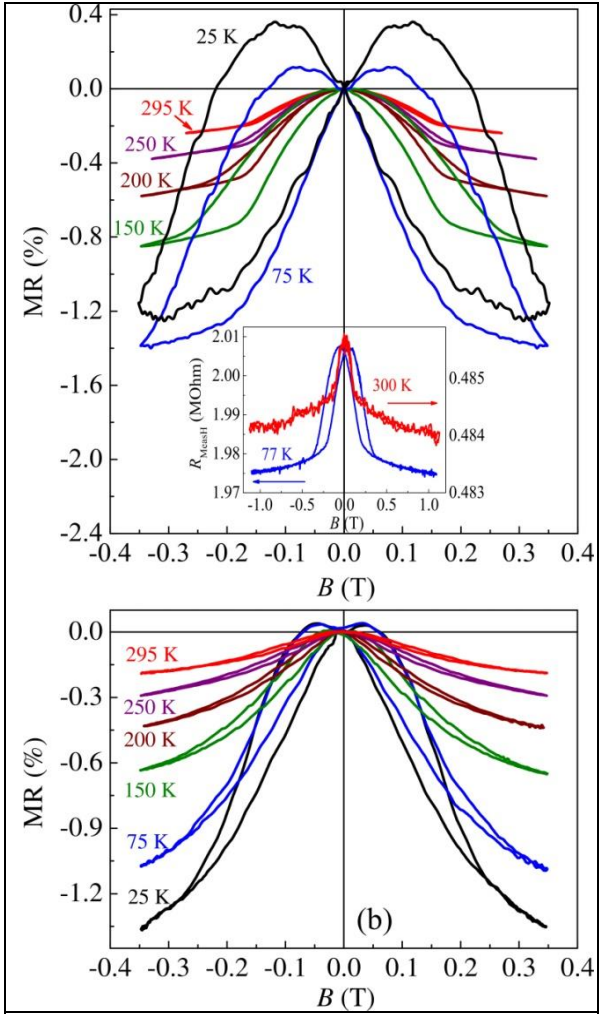

FIG. 12. MR curves at various temperatures for structure 200-25. (a) $B$ is applied perpendicular to sample plane (inset shows $R_{MeasH}(B)$ curves for $B = \pm 1.1$ T at 77 and 300 K). (b) $B$ is applied in sample plane.

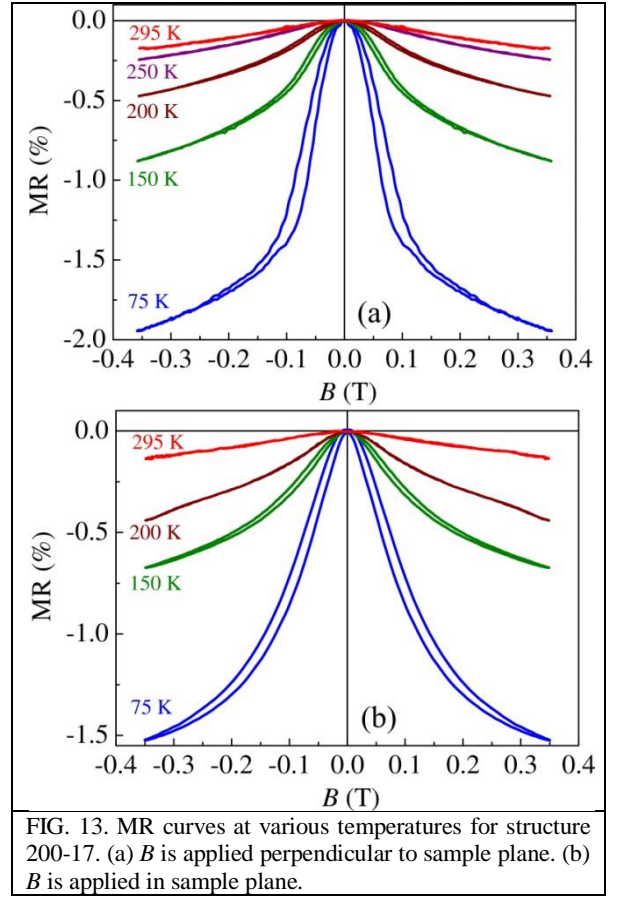

FIG. 13. MR curves at various temperatures for structure 200-17. (a) $B$ is applied perpendicular to sample plane. (b) $B$ is applied in sample plane.

Figure 12 shows the out-of-plane (Fig. 12(a)) and in-plane (Fig. 12(b)) MR curves for the sample 200-25 in the temperature range of 25 – 295 K. For the structure 200-25 the magnetoresistance is negative and hysteretic up to 295 K. A comparison of the out-of-plane (Fig. 2n temperature also points to the out-of-plane magnetic anisotropy of the GaAs:Fe layer of the structure 200-25. Inasmuch as the MR curves are hysteretic up to 295 K the $T_C$ for the structure 200-25 is not less than room temperature.

Figure 13(a) shows the MR curves for the sample 200-17 in the temperature range of 75 – 295 K with $B$ applied perpendicular to the plane. As for the structure 200-25 the magnetoresistance is negative up to 295 K but the hysteresis disappears at temperatures above 200 K. The out-of-plane MR curves also show that the coercivity of the GaAs:Fe layer for the structure 200-17 is smaller than that for the structure 200-25 (Figure 12) in the all temperature range. The Fe concentration decrease in the structure 200-17 in comparison with the structure 200-25 leads to the weakening of the ferromagnetic properties of the GaAs:Fe layer and to $T_C$ reduction (down to ~ 200 K). The MR investigations also reveal the out-of-plane magnetic anisotropy of the GaAs:Fe layer for the structure 200-17 as for the structures 180-25 and 200-25.

For the structure 200-8 no reliable MR curves were obtained due to the higher resistance of the GaAs:Fe layer with the lowest Fe content.

To confirm that the ferromagnetic properties of the obtained layers are related to the intrinsic ferromagnetism of GaAs:Fe DMS, the studies of a magnetic circular dichroism were carried out for the structures 180-25, 200-25, 250-25 and 350-25. The MCD investigations were performed for $B$ applied perpendicular to the sample plane. The light from a xenon arc lamp, after passing through a grating monochromator, was circularly polarized by the combination of a linear polarizer and an achromatic quarter-wave plate. This light was incident on the sample surface and reflected from it into the measuring system (fotodetector). The MCD value was defined as $((I^+ - I^-)/(I^+ + I^-)) \times 100\%$, where $I^+$ and $I^-$ are the intensities of circularly polarized light with the right and left circular polarization incident on the sample's surface.

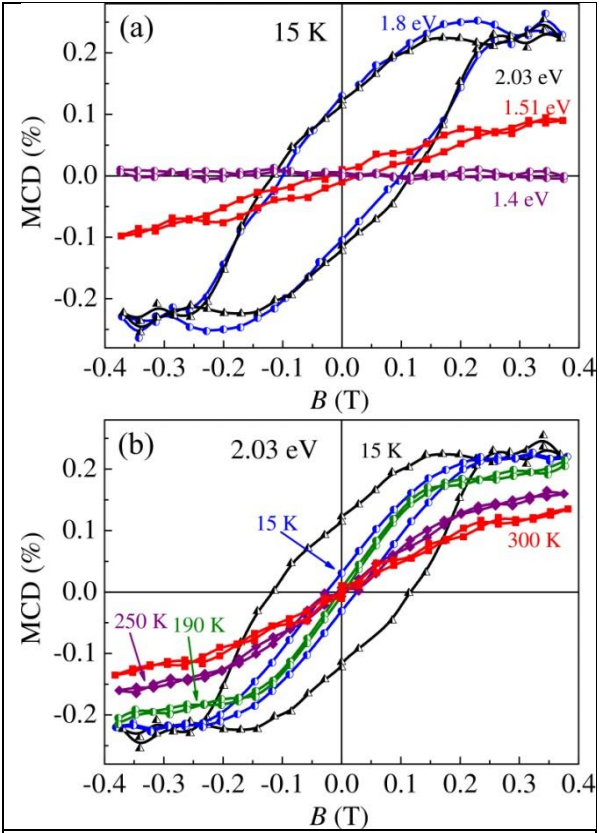

FIG. 14. (a) MCD($B$) dependences for structure 180-25 measured for different photon energies at 15 K. (b) MCD($B$) dependences for structure 180-25 at different temperatures for photon energy of 2.03 eV.

Figure 14 shows the MCD dependences on an external magnetic field (MCD($B$)) for the structure 180-25 measured at 15 K for the photon energies in the range of 1.4 – 2.03 eV. For the photon energies of 2.03 and 1.8 eV that are notably higher than the GaAs bang gap ($E_g$) at 15 K (~ 1.52 eV), the MCD($B$) dependences are hysteretic with the saturation at $B > 0.3$ T and have the maximum amplitude (Fig. 14(a)). When the photon energy decreases to 1.51 eV the amplitude of the MCD($B$) dependence drops greatly. For the photon energy of 1.4 eV that is smaller than $E_g$ of GaAs, the detectable MCD value is practically absent (Fig. 14(a)). Figure 14(b) shows the MCD($B$) curves for the structure 180-25 for the photon energy of 2.03 eV in the temperature range of 15 – 300 K. The MCD($B$) investigations at different temperatures reveal (using the Arrot plot of the MCD($B$) curves) that the Curie point for the structure 180-25 is about 200 K, this is in agreement with the result of magnetoresistance investigations (Fig. 11).

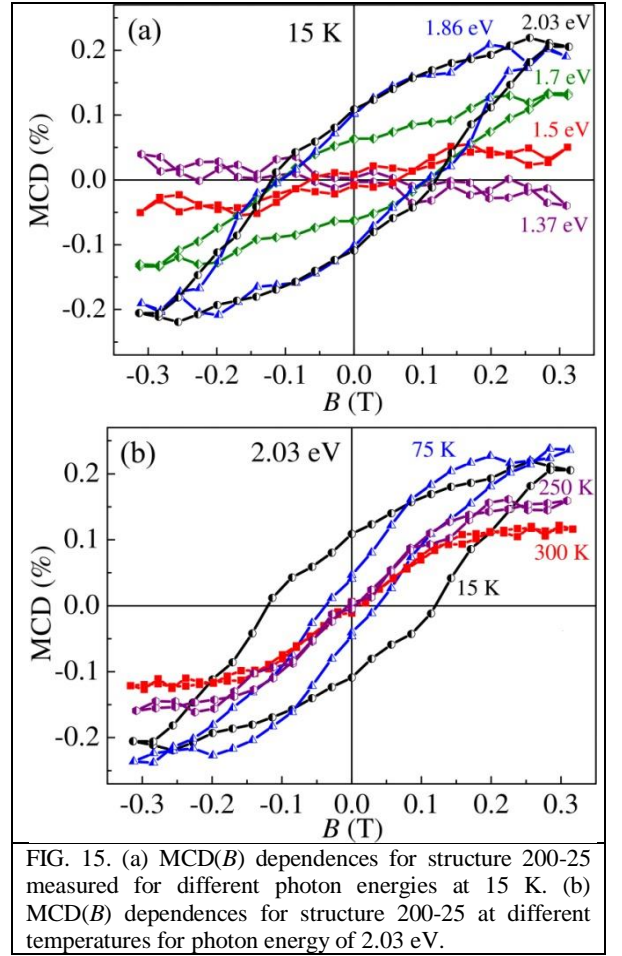

FIG. 15. (a) MCD($B$) dependences for structure 200-25 measured for different photon energies at 15 K. (b) MCD($B$) dependences for structure 200-25 at different temperatures for photon energy of 2.03 eV.

Figure 15(a) exhibits the MCD($B$) dependences at 15 K for the structure 200-25 measured for the photon energies in the range of 1.37 – 2.03 eV. While the photon energies are higher than the GaAs bang gap, the MCD($B$) dependences are hysteretic without the saturation. Note that the hysteretic unsaturated shape of the MCD($B$) curves at 15 K is completely consistent with the out-of-plane MR curve at 25 K (Fig. 12(a)). Similarly to as it was observed for the structure 180-25 (Fig. 12(a)), for the photon energy of 1.5 eV that is smaller than $E_g$ of GaAs, the MCD amplitude drops greatly and for the photon energy of 1.37 eV the detectable MCD effect is practically absent (Fig, 15(a)). Figure 15(b) shows the MCD($B$) dependences for the structure 200-25 for the photon energy of 2.03 eV in the temperature range of 15 – 300 K. The MCD($B$) curves confirm that the GaAs:Fe layer of the structure 200-25 is ferromagnetic up to RT. The absence of the detectable MCD value for the photon energies ≤ 1.5 eV (when GaAs becomes transparent) reveal that the observed ferromagnetic MCD($B$) curves at 15 K (at the photon energies ≥ 1.7 eV) are related to the intrinsic ferromagnetism of the GaAs:Fe layers but not to any second-phase of Fe or intermetallic ferromagnetic inclusions. The spectral dependence of MCD also shows that the $E_g$ of obtained GaAs:Fe layers is close to $E_g$ of GaAs. Thus the MCD investigations confirm that the obtained conductive GaAs:Fe layers of the structures 180-25 and 200-25 is diluted magnetic semiconductor with intrinsic ferromagnetic properties up to room temperature for the structure 200-25. Since the GaAs:Fe layers apparently contain DMS regions with the Fe concentration above average (periodic columnar-shape domains), separated by DMS regions with the Fe concentration below average, these regions have different Curie temperature. The high-temperature intrinsic ferromagnetism evidently is associated with the columnar Fe-enriched DMS regions.

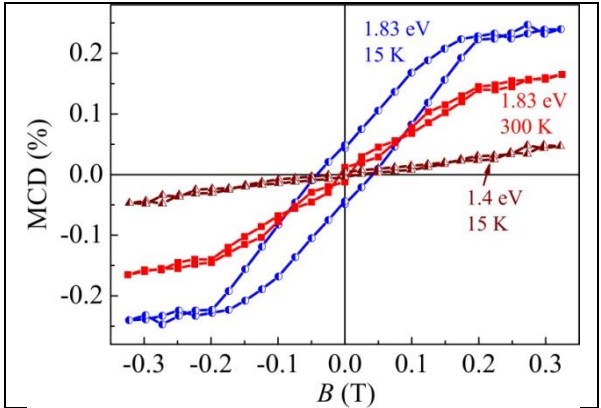

FIG. 16. MCD(*B*) dependences for structure 250-25 measured at 15 K for photon energies of 1.83 and 1.4 eV and measured at 300 K for photon energy of 1.83 eV.

The similar character of the MCD effect is also observed for the structure 250–25. Figure 16 exhibits the MCD(*B*) dependences for the structure 250-25 measured at 15 K for the photon energies of 1.83 and 1.4 eV and measured at 300 K for the photon energy of 1.83 eV. At 15 K and the photon energy of 1.83 eV the MCD(*B*) dependence is clearly hysteretic while for the photon energy of 1.4 eV the MCD(*B*) dependence is weak. The observed threshold of the MCD effect close to $E_g$ of GaAs confirms that the clusters observed in the structure 250–25 can be associated with the Fe-rich regions of GaAs:Fe DMS (Fig. 6).

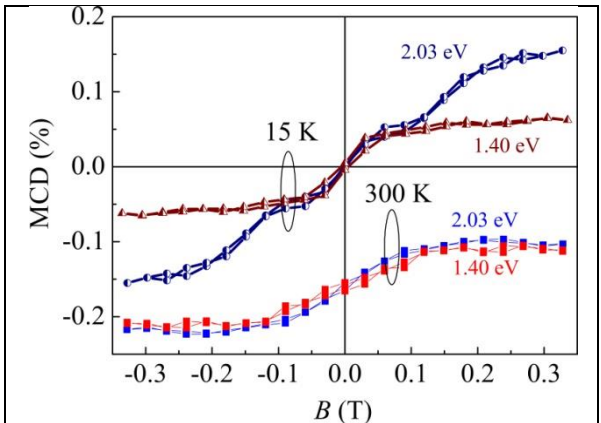

FIG. 17. MCD(*B*) dependences for structure 350-25 measured at 15 and 300 K for photon energies of 2.03 and 1.4 eV.

A fundamentally different character of the MCD effect is observed for the structure 350–25. Figure 17 shows the MCD(*B*) dependences for the structure 350-25 measured at 15 and 300 K for the photon energies of 2.03 and 1.4 eV. The MCD investigations show that for this structure there is a clear nonlinear contribution to the MCD(*B*) curves independent of the photon energy and the temperature. Consequently, the growth temperature increase to 350 ºC leads to the formation of metallic ferromagnetic second-phase inclusions, presumably due to the coalescence of Fe atoms similar to that was observed in the (In,Fe)Sb layer obtained at 300 ºC [17].

## 4. Discussion

The fabricated GaAs:Fe layers have the rather high resistivity but for the structures 180-25, 200–25 and 200–17 the conductivity is also observed at low temperatures (Fig. 10). This result is interesting in by itself, since Fe atoms are located at substitutional Ga sites ($Fe_{Ga}$) forms deep acceptor levels in the GaAs band gap (about 0.88 and 0.51 eV above the valence band (VB) edge [29-33]) and thus should be a compensating impurity. For the Fe atom in the Ga position, two valence *s* electrons and one *d* electron participate in the formation of chemical bonds with As atoms [33]. In this case the $Fe_{Ga}$ atom can be in two different valence states: $Fe^{3+}$ ($3d^5$) state - as a neutral impurity (neutral acceptor) and $Fe^{2+}$ ($3d^6$) state - as an ionized deep acceptor [33,34].

Based on the results of our XPS studies (Fig. 8, Fig. 9) it can be concluded that in the fabricated GaAs:Fe layers the significant number of Fe atoms (up to 50 %) are located at substitutional As sites ($Fe_{As}$). At first glance, this inference is controversial. As shown in [35], the most favorable position of Fe in GaAs is at Ga site, while the solution energy of Fe at the interstitials is less than in $Fe_{As}$ position. The difference between the formation energies of $Fe_{As}$ and $Fe_I$, where *I* stands for ″interstitial″ is about 1.5 eV depending on type of interstitial [35]. Meanwhile, a large number of $Fe_{Ga}$ defects should lead to appearance of $Ga_I$ or $Ga_{As}$ because of observed almost equal number of Ga and As atoms in the sample. The formation energy of $Ga_I$ is about 4.5 eV [35, 36] and $Ga_{As}$ antisite is about 3.0 eV [35, 37]. These values exceed the difference in formation energies of $Fe_{As}$ and $Fe_I$. Thus, solution of Fe into As sublattice can be more preferable than formation of Fe interstitials together with accompanying defects like $Ga_I$ or $Ga_{As}$. No information in the literature about the possible character of the electrical activity of $Fe_{As}$ atoms were found. Since Fe atom has only two valence 4*s* electrons and has the acceptor nature in the position of the group III element, it is most likely that the Fe atom in the group V position also acts as an acceptor (deep acceptor, taking into account the high resistivity of the layers). The acceptor nature of $Fe_{As}$ atoms should also be related to the different occupation of the *d* shell after the electron accepting. Comparing Ga and As atomic configurations and taking into account mechanism of $Fe^{3+}$ acceptor formation in Ga site with hybridization of Fe *d* shells and As *p* orbitals [38], one can draw a preliminary conclusion that $Fe_{As}$ can be deep acceptor with the $Fe^{5+}$ state, that differs from the $Fe^{3+}$ state of $Fe_{Ga}$ center.

It is known, that for the high concentration of an impurity the formation of an impurity band (IB) is possible. In particular, 3*d* elements such as Mn and Ni form the acceptor IB in GaAs [39]. The ferromagnetic properties and the current transfer in the (Ga,Mn)As DMS can be determined by charge carriers in the Mn impurity band separated from VB by a finite energy gap [40-42]. The investigations of the spatial structure of the Fe wave function in GaAs by a scanning tunneling microscopy reveal that the spatial expansion of the Fe wave function is relatively large and can reach 2 – 2.5 nm in different directions [33,43]. In our GaAs:Fe layers, the Fe concentration is very high, the Fe local concentration can exceed 25 at. % (Fig. 5) and in this case the average distance between neighboring Fe atoms is less than spatial expansion of the Fe wave function which makes it reasonable to assume that the Fe deep impurity (both $Fe_{Ga}$ and $Fe_{As}$) can form the acceptor IB within GaAs band gap. It can be assumed, that the observed transport properties of the obtained GaAs:Fe conductive layers are determined by the electron transport in the Fe-related IB. In particular, this is indicated by the weak temperature dependence of the resistivity (Fig. 10).

The mobility of the carriers localized within IB should be small (about or less than 1 $cm^2$/V·s). Also, as it was shown by TEM, STEM and EDS studies, the periodic vertical structure with difference in the Fe concentration is observed within the conductive GaAs:Fe layers of the structures 180-25 and 200-25 (Figures 2 – 5). This should lead to the spatial inhomogeneity of the impurity band, to the strong modulation of the built-in electric field and consequently to the strong modulation the potential profile. These phenomena should further reduce the effective carrier mobility which finally leads to the observed high resistivity of the conductive structures.

Taking into account the effective mobility in the range of about 0.1 - 1 $cm^2$/V·s and the resistivity of 0.5 Ohm·cm (the resistivity at 300 K for the structure 180-25, Fig. 10), the carrier concentration of an order of magnitude of $10^{19}$ - $10^{20}$ $cm^{-3}$ can be obtained. This is much lower than the concentration of introduced Fe atoms. But this is possible, since the Fe acceptor impurity forms the deep (originally empty) acceptor IB in which the charge carriers (electrons) are delivered from some additional donor levels (in particular point defects) and the resulting concentration of electrons participating in the conductivity can be much lower than the acceptor (Fe) concentration. It is most likely that for the conductive structures 200–8, 200–17, and 200–25 the carrier concentration also does not exceed $10^{20}$ $cm^{-3}$.

The relatively low carrier concentration indicates that, as in other Fe doped III-V DMSs, the ferromagnetism in the GaAs:Fe system apparently is not carrier-mediated. As was mentioned above, the relatively large spatial expansion of the Fe wave function and the high Fe concentration allow the overlapping of the wave functions of neighboring Fe atoms which permits the direct exchange interaction between Fe atoms.

The feature of the GaAs:Fe system is also that Fe atoms act as an electrically active impurity (at least $Fe_{Ga}$ atoms) and can be in two valence states with the different occupation of the *d* shell (the neutral and ionized acceptor states). Since the nearest Fe atoms are separated by the As and Ga atom (with which Fe atoms have a bond via overlapping the *d* (Fe) and *p* (As and Ga) shells), the electron transfer between the acceptor states with different valency is possible which allows the realization of ferromagnetic ordering via the Zener double exchange mechanism [44]. This mechanism of ferromagnetic exchange interaction in GaAs:Fe seems more preferable. More detailed study of the exchange interaction mechanisms in the GaAs: Fe system requires the involvement of *ab initio* theoretical research. Our presented experimental results definitely show that ferromagnetism (including at RT) may occur in the heavily Fe-doped GaAs:Fe DMS layers.

## 5. Conclusion

In summary, conductive *n*-type GaAs:Fe layers with an average Fe content up to 20 at. % were obtained by the PLD. The TEM, STEM and EDS investigations of the conductive GaAs:Fe layers with the maximum Fe content obtained at 180 and 200 ºC revealed that the layers are epitaxial, do not contain any second-phase inclusions, but contains the columnar Fe-enriched regions of overlapped microtwins. These layers can be characterized as a vertical periodic system consisting of nano-columnar regions of a GaAs:Fe DMS with an above-average Fe concentration separated by regions of a GaAs:Fe DMS with a below-average Fe concentration, similar to what was previously observed for MBE (Ga,Fe)Sb layers [4]. The TEM investigations of the non-conductive GaAs:Fe layer the maximum Fe content obtained at 250 ºC revealed that the layer is the epitaxial GaAs:Fe DMS matrix with the embedded coherent Fe-rich clusters of GaAs:Fe DMS. The XPS investigations showed that Fe atoms form chemical bonds with Ga and As atoms with almost equal probability, therefore, the comparable number of Fe atoms substitute on Ga and As sites. Presumably this makes it possible to obtain a very high concentration of Fe atoms in the LT GaAs:Fe layers significantly exceeding the equilibrium solubility limit (about $10^{18}$ $cm^{-3}$, [23, 45]) without formation of metallic Fe or intermetallic second-phase clusters, in particular, $Fe_3GaAs$ [23].

The conductivity of the GaAs:Fe layers is apparently associated with the formation of the Fe acceptor impurity band within the GaAs band gap. A hysteretic negative magnetoresistance was observed in the GaAs:Fe layers up to room temperature. Magnetoresistance investigations point to the out-of-plane magnetic anisotropy of the GaAs:Fe layers which is related to the presence of the columnar Fe-enriched GaAs:Fe regions. The investigations of the magnetic circular dichroism confirm that the GaAs:Fe layers with the maximum Fe content obtained at 180, 200 and 250 °C are intrinsic ferromagnetic semiconductors with the Curie point up to at least room temperature for the conductive layer obtained at 200 °C. Heavily Fe-doped doped (up to ~ 20 at. %) GaAs layers are resistant to the formation of metallic ferromagnetic second-phase ferromagnetic inclusions for the growth temperatures up to about 250 °C. The growth temperature increase up to 350 °C leads to the formation of metallic ferromagnetic second-phase inclusions.

It was suggested that in heavily Fe-doped GaAs DMS layers the ferromagnetism is related to the Zener double exchange between Fe atoms with different valence states (the neutral and ionized acceptor states) via an intermediate As and Ga atoms. Thus, the possibility of synthesis of conductive GaAs:Fe DMS layers with the intrinsic ferromagnetic properties and high Curie temperature has been shown.

**Acknowledgements**

This study was supported by grant № 18-79-10088 from Russian Science Foundation. N.A Sobolev and S.Yu. Zubkov are gratefully acknowledged for discussion and for the critical reading of the manuscript.